\documentclass[final,3p,times,twocolumn]{elsarticle}

\usepackage[T1]{fontenc}
\usepackage[utf8]{inputenc}
\usepackage{amsmath}        
\usepackage{epstopdf}     
\usepackage{flushend}         
\usepackage{hyperref}        
\usepackage{graphicx}
\usepackage{lineno}
\usepackage{ulem}
\usepackage{miller}

\newcommand{\etal}{\textit{et al}}
\newcommand{\ie}{\textit{i.e.}}
\newcommand{\eg}{\textit{e.g.}}

\newcommand{\abinitio}{\textit{ab initio}}
\newcommand{\Abinitio}{\textit{Ab initio}}
\newcommand{\insitu}{\textit{in situ}}

\usepackage{xcolor}

\journal{Acta Materialia}

\begin{document}

\begin{frontmatter}


\title{Screw dislocation-carbon interaction in BCC tungsten: an \abinitio{} study}

\author[SRMP]{Guillaume Hachet\corref{CA}}
\ead{guillaume.hachet@cea.fr}
\author[SRMP]{Lisa Ventelon}
\author[SRMP]{François Willaime}
\author[SRMP]{Emmanuel Clouet\corref{CA}}
\ead{emmanuel.clouet@cea.fr}
\cortext[CA]{Corresponding authors}

\address[SRMP]{Université Paris-Saclay, CEA, Service de Recherches de Métallurgie Physique, 91191, Gif-sur-Yvette, France}

\begin{abstract}
The interaction between carbon and screw dislocations in tungsten is investigated using \abinitio{} calculations. 
The presence of carbon atoms in the vicinity of the dislocation induces a reconstruction, with the dislocation relaxing to a configuration, the hard core structure, which is unstable in pure tungsten.
The reconstruction corresponds to a strong binding of carbon in the prismatic sites created by the dislocation which is perfect for high concentrations of carbon segregated on the dislocation line. 
However, the reconstruction is only partial for lower atomic fractions, with the dislocation tending to fall back in its easy core ground state.
This pinning by carbon atoms of the dislocation in an unstable position is well described by a simple line tension model.
A strong carbon-dislocation attraction is also evidenced at larger separation distances, when the solute is in the fourth nearest neighbour octahedral sites of the reconstructed core.
The equilibrium concentrations of carbon in these different segregation sites are modelled with an Ising model and using a mean-field approximation.
This thermodynamic model evidences that screw dislocations remain fully saturated by carbon atoms and pinned in their hard core configuration up to about 2500\,K.
\end{abstract}


\end{frontmatter}


\section{Introduction}
\label{S1}

Plasticity of body centred cubic (BCC) tungsten, like other BCC metals, is mainly controlled by the motion of dislocations with $1/2\,\hkl<111>$ Burgers vectors \cite{Hirth1982,Caillard2003}. 
At low temperatures, below $\sim 0.2\,T_m$ ($T_m$ is the melting temperature), the strong temperature and strain-rate dependence of the flow stress \cite{Argon1966,Brunner2000,Brunner2010} derives from the thermally activated glide of the screw orientation.
Transmission electron microscopy (TEM) on strained tungsten single crystals \cite{Stephens1970} shows that dislocations preferentially align with their screw orientation, an indication of a high lattice friction opposing the glide of screw dislocations. 
Indeed, these dislocations usually glide by the so-called Peierls mechanism, which results in a steady and smooth motion as confirmed by \textit{in situ} TEM straining experiments \cite{Caillard2018}.
The $1/2\,\hkl<111>$ screw dislocations appear therefore as the rate controlling mechanism of plasticity in BCC tungsten.

Interstitial solutes like carbon atoms strongly modify the mechanical properties of BCC metals.
Tensile tests on single and polycrystalline tungsten have demonstrated an increase of the yield stress with the carbon content \cite{Stephens1964}.
Such a hardening is associated with a pinning of gliding dislocations by carbon in solid solution. 
This is confirmed by internal friction measurements, which show, in addition to the classical Snoek peak, a second peak (the Köster peak), corresponding to the interplay between dislocations and carbon atoms \cite{Carpenter1965,Gray1971}.
The seminal work of Cottrell and Bilby \cite{Cottrell1949} has demonstrated that such a pinning of dislocations can be well understood by their elastic interactions with solute atoms, even for screw dislocations when the non-dilatational part of the strain caused by the interstitial foreign atom is considered \cite{Cochardt1955}.
However, although elasticity theory perfectly accounts for this interaction when the dislocation and the solute atom are a few atomic distances apart \cite{Clouet2008}, an atomic description is needed for configurations potentially leading to the strongest pinning, \ie{} when the carbon atom is in the immediate vicinity of the dislocation core.
In order to better understand the effects of carbon atoms in solid solution on tungsten plasticity, atomistic simulations of their interactions with screw dislocations appear therefore as a valuable and essential tool.

\Abinitio{} calculations have shown that the $1/2\,\hkl<111>$ screw dislocation has a compact core in BCC tungsten \cite{Romaner2010,Ventelon2013,Weinberger2013,Dezerald2014}, like in 
other BCC metals \cite{Woodward2002,Rodney2017}.
The glide of this screw dislocation accounts for the temperature dependence of the yield stress \cite{Dezerald2015} and for the deviation from Schmid law \cite{Dezerald2016,Kraych2019} in pure tungsten.
Considering the interaction of this dislocation with different solute atoms \cite{Romaner2010,Li2012,Itakura2012,Itakura2013,Hu2017,Tsuru2018,Zhao2020,Luthi2017,Bakaev2019}, these \abinitio{} calculations have also shed some new light on the physical mechanisms controlling solute hardening or softening in tungsten.
More specifically in the case of carbon, a strong attraction between the dislocation and the solute atom has been evidenced \cite{Luthi2017,Bakaev2019}.
This attractive interaction induces a spontaneous reconstruction of the dislocation core towards a configuration, called hard core structure, which is strongly unstable in pure tungsten \cite{Dezerald2014}.
The resulting regular prism formed by tungsten atoms is stabilised by the segregation of carbon at its centre, thus pinning the dislocation.
Such a core reconstruction induced by carbon is not specific to tungsten: it has been first found in iron \cite{Ventelon2015,Luthi2019} and appears to be generic to BCC transition metals \cite{Luthi2017}.
This core reconstruction towards the hard core configuration was also evidenced when the screw dislocation interacts with other interstitial solute elements \cite{Luthi2018}, except hydrogen which induces a different reconstruction towards the split-core configuration, centred in the immediate vicinity of a $\hkl<111>$ atomic column \cite{Grigorev2020}.

The aim of the present article is to investigate this dislocation core reconstruction induced by carbon and the resulting carbon segregation on screw dislocation in tungsten. 
\Abinitio{} calculations are first performed to describe the locking of the screw dislocation in its hard core configuration for different atomic fractions of carbon segregated in the dislocation core.
Results are used to develop a line tension model, which allows the extrapolation to the dilute case where a long screw dislocation interacts with a single carbon atom. 
We then consider other possible attractive sites for carbon in the immediate vicinity of the dislocation core and build a thermodynamic model of carbon segregation around the dislocation line as a function of temperature.

\section{Methods}
\label{S2}

\subsection{Computational details}
\label{S21}

We perform \abinitio{} calculations using density functional theory (DFT) implemented in the Vienna \abinitio{} simulation package (VASP) code \cite{Hohenberg1964,Kohn1965,Kresse1996}.
Pseudopotentials built with the projected augmented wave method \cite{Blochl1994,Kresse1999} are used with a kinetic energy cutoff of 400\,eV. 
Only 5d and 6s electrons for tungsten and 2s and 2p electrons for carbon are included in valence states.
We have checked that the inclusion of semi-core electrons in tungsten pseudopotential does not modify the dislocation-carbon interaction energy and have therefore chosen to omit these electrons.
The exchange-correlation is described with the generalised gradient approximation with the Perdew-Burke-Ernzerhof functional \cite{Perdew1996}.
All calculations are performed at constant cell volume with a 0.2\,eV Hermite Gaussian broadening.
For the $1\,b$ dislocation simulation cell described below ($b$ is the dislocation 
Burgers vector), we use a 2$\times$2$\times$16 shifted $k$-point grid to sample the Brillouin zone and the numbers of $k$-points are 8, 6, 4, 4, 3, 2 and 2 along the Z reciprocal directions 
respectively for cell heights $h$ = 2, 3, 4, 5, 6, 8 and 10 \,b.
Calculations in the perfect BCC crystal without dislocation are performed in a 250-atoms cubic cell using a 4$\times$4$\times$4 shifted $k$-point grid.
Atomic positions are relaxed until all ionic forces are inferior to $10^{-2}$\,eV/\AA.
With these parameters, the obtained lattice parameter is equal to 3.173\,\AA{} and elastic constants are $C_{11}$ = 497 GPa, $C_{12}$ = 227 GPa and $C_{44}$ = 131 GPa.

\subsection{Simulation cell setup}
\label{S22}

All \abinitio{} calculations of dislocations are performed in a periodic supercell containing a dislocation dipole leading to a quadrupolar periodic array of dislocations \cite{Rodney2017,Clouet2018}. 
The periodicity vectors $\{\mathbf{p}_1,\mathbf{p}_2,\mathbf{p}_3\}$ of the perfect supercell are defined from the elementary vectors $\mathbf{u}_1=\hkl[-1-12]$, $\mathbf{u}_2=\hkl[1-10]$  
and $\mathbf{u}_3=1/2\,\hkl[111]$: 
$\mathbf{p}_1 = 5/2\,\mathbf{u}_1-9/2\,\mathbf{u}_2$,
$\mathbf{p}_2 = 5/2\,\mathbf{u}_1+9/2\,\mathbf{u}_2$ 
and $\mathbf{p}_3 = \mathbf{u}_3$.
Such a configuration implies 135 tungsten atoms per $1\,b$ layer along the $Z$ axis in the \hkl[111] direction parallel to the dislocation line \cite{Ventelon2013} and the distance between 
dislocations is 19\,\AA{} (see figure in supplementary materials for an illustration of the simulation cell). 
Previous studies have shown that such simulation cell using a quadrupolar arrangement is large enough to obtain well-converged energies for dislocation cores \cite{Clouet2009a,Ventelon2013}.
Then, this supercell is duplicated up to ten times along the $Z$ axis to model various carbon concentrations on the dislocation line.
With this setup, we obtain an energy difference $\Delta E_{H-E}$ between the hard core configuration of the screw dislocation and its easy core ground state in pure tungsten of 129\,meV/$b$, in  
good agreement with previous DFT calculations \cite{Dezerald2014,Luthi2017}.

The same number of carbon atoms is inserted in the cores of both dislocations forming the dipole at the same relative position.  
We check that the dislocations and their carbon atoms relax to the same configuration. 
The interaction energy between the carbon atoms and the screw dislocation is then defined by:
\begin{multline}
\label{eqEintTot}
E^{\textrm{inter}}_{\textrm{D-}n\textrm{C}} = \frac{1}{2} \left[  E(\textrm{D}+n\,\textrm{C}) - E(\textrm{D})\right] \\ 
	- n \left[ E(\textrm{C}) - E(\textrm{bulk}) \right],
\end{multline}
with $n$ the number of carbon atoms on each dislocation.
$E(\textrm{D}+n\,\textrm{C})$ and $E(\textrm{D})$ are the energies of the same supercell containing respectively both the dislocation dipole and the carbon atoms, and only the dislocation dipole with 
dislocations in their ground state.
$E(\textrm{C})$ and $E(\textrm{bulk})$ are the energies of the same BCC supercell with and without a carbon atom in its stable octahedral position.
With such a definition, attraction between the carbon atoms and the screw dislocation leads to a negative interaction energy.

When several carbon atoms are inserted in the vicinity of a screw dislocation (\S \ref{S41}), we define an incremental interaction energy:
\begin{equation}
	\Delta E^{\rm inter}_{\textrm{D-C}} 
		= E^{\rm inter}_{\textrm{D-}n\textrm{C}}
		- E^{\rm inter}_{\textrm{D-}(n-1)\textrm{C}}.
	\label{eqEintInc}
\end{equation}
It describes the interaction energy of the last arriving C atom with the already populated dislocation.

\section{Carbon atoms in the dislocation core}
\label{S3}

We first examine the interaction of a screw dislocation with a single $\hkl<111>$ column of carbon atoms. 
We vary the occupation $x_{\rm C}$ of the carbon atomic column by using different heights $h$ along $\hkl<111>$ ranging from $h=1\,b$ (135 tungsten atoms) corresponding to a fully occupied column,  
$x_{\rm C}=1$, to $h=10\,b$ (1350 tungsten atoms) for the lowest carbon atomic fraction $x_{\rm C}=0.1$.

\subsection{Dislocation core reconstruction}
\label{S31}

\begin{figure}[bth]
\centering
\includegraphics[width=0.99\linewidth]{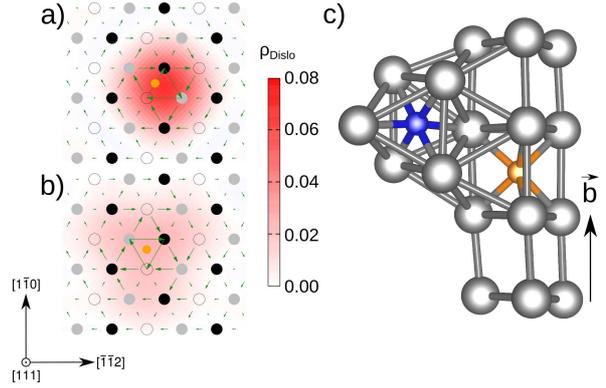}
\caption{Core reconstruction of a screw dislocation in the presence of a carbon atom for $h=1\,b$ (a) before and (b) after atomic relaxation.
In these projections perpendicular to the dislocation line, tungsten atomic columns are sketched by symbols with a colour depending on their (111) plane in the original perfect crystal.
The arrows between two atomic columns are proportional to the differential displacement created by the dislocation in the [111] direction. 
The contour map shows the dislocation density according to the Nye tensor.
The carbon atom in the vicinity of the dislocation core is shown in orange.
(c) 3D visualisation of the dislocation in hard core configuration with one carbon in the prismatic site (orange) and one carbon in the fourth nearest neighbour octahedral site (blue) of the  
reconstructed core studied in \S \ref{S4}.}
\label{figVMNT1b}
\end{figure}

Depending on the local atomic arrangement, the dislocation core can have different configurations in pure BCC metals.
Its ground state \cite{Vitek1974,Dezerald2014,Rodney2017}, called easy core, leads locally to the inversion of the helicity of the BCC lattice, 
with a reversed order of the heights of the three \hkl[111] atomic columns defining the core.
A second configuration, called hard core, where the three \hkl[111] atomic columns are at the same height, is unstable in pure BCC crystals \cite{Vitek1974,Dezerald2014,Rodney2017}
but can be stabilised by interstitial solute atoms \cite{Ventelon2015,Luthi2017,Luthi2018,Luthi2019,Bakaev2019}.
Considering first a dislocation in its easy core ground state interacting with a carbon atom in a first nearest neighbour octahedral site (Fig. \ref{figVMNT1b}a), we observe a systematic dislocation core reconstruction from the easy to the hard core configuration.
The dislocation centre is shifted to the neighbouring triangle defined by $\hkl[111]$ tungsten atomic columns and the solute atom moves in the newly created prismatic interstitial site (Fig. \ref{figVMNT1b}b).
This structure with regular trigonal prisms formed by the metal atoms with a carbon at their centre is similar to the unit building blocks of hexagonal WC tungsten carbide (Fig \ref{figVMNT1b}c) 
\cite{Luthi2017}.

Previous \abinitio{} studies \cite{Luthi2017,Bakaev2019} have evidenced the same core reconstruction.
It is observed in various BCC transition metals when the screw dislocation interacts with interstitial solute atoms like C, B, N, or O \cite{Ventelon2015,Luthi2017,Luthi2018,Luthi2019,Bakaev2019}.

\subsection{Stabilisation of the hard core configuration}
\label{S32}

\begin{figure}[!tbh]
\centering
\includegraphics[width=0.99\linewidth]{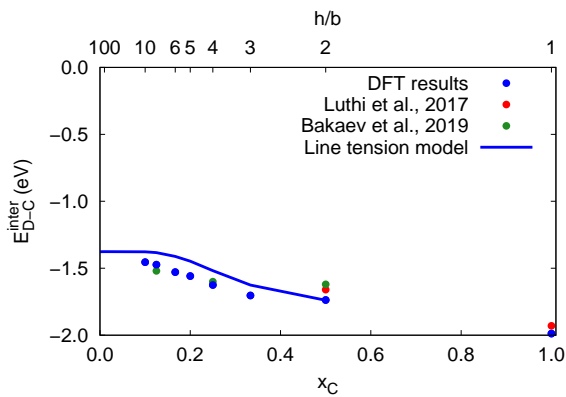}
\caption{Interaction energy $E^{\rm inter}_{\rm D-C}$ as a function of the fraction $x_{\rm C}$ of carbon atoms in the dislocation core or equivalently of the distance $h$ between carbon atoms along the dislocation line. 
The symbols correspond to \abinitio{} calculations, including previous work of L\"{u}thi \etal{} \cite{Luthi2017} and Bakaev \etal{} \cite{Bakaev2019}, and the solid line to the line tension model (Eq. \ref{eqModel}).}
\label{figEinter}
\end{figure}

As the screw dislocation reconstructs towards its hard core configuration in the presence of carbon, we have begun directly from the hard core configuration and inserted the solute in the prismatic site for various supercell heights $h$.
For all corresponding atomic fractions ($0.1 \leq x_{\rm C} \leq 1$), after relaxation, the carbon atom stays in its initial position and locally pins the dislocation in its hard core configuration.
The obtained interaction energy $E^{\rm inter}_{\rm D-C}$ (Eq. \ref{eqEintTot}) is shown in figure \ref{figEinter}. 
The interaction is strongly attractive: the fully saturated dislocation line ($x_{\rm C}=1$) is the most attractive configuration for which $E^{\rm inter}_{\rm D-C}=-1.99$\,eV.
Then, the interaction energy decreases in absolute value when less carbon atoms are present on the dislocation line but remains still large, $E^{\rm inter}_{\rm D-C}=-1.45$\,eV, for the most dilute 
case $x_{\rm C}=0.1$ considered in this study.
The obtained interaction energy agrees with previous DFT studies \cite{Luthi2017,Bakaev2019}.
Moreover, we notice that $E^{\rm inter}_{\rm D-C} (x_{\rm C} = 1) < E^{\rm inter}_{\rm D-C} (x_{\rm C} = 0.5) $, meaning that the nearest neighbour carbon-carbon interaction is attractive along the dislocation line.
This contrasts with iron, where this carbon-carbon interaction is repulsive \cite{Ventelon2015}. 
However, this result is consistent with the carbon-carbon interaction energy calculated in the dislocation-free tungsten crystal. 
Indeed this energy for a distance of $1\,b$ between carbon atoms along $\hkl<111>$ in the bulk is about $-0.4$\,eV \cite{Luthi2017}, which is even more attractive than along the dislocation line where this pair interaction is equal to $-0.08$\,eV as it will be shown in section \ref{S42}.

\begin{figure}[htbp]
\centering
\includegraphics[width=0.99\linewidth]{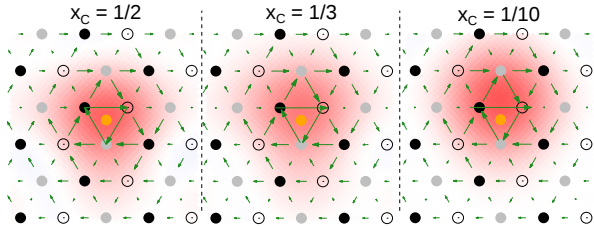}
\caption{Dislocation core structure in the $b$-layer containing the carbon atom for different atomic fractions $x_{\rm C}$ of carbon (see caption of Fig. \ref{figVMNT1b} for a detailed description).}
\label{figDispMaps}
\end{figure}

We now examine in more details the atomic structure of the screw dislocation with the carbon atom in the prismatic site.
For a high concentration of segregated carbon ($x_{\rm C}\geq0.5$), the dislocation keeps a structure corresponding to a perfect hard core configuration (Fig. \ref{figDispMaps}). 
For lower atomic fractions of carbon, the structure does not exactly correspond to this hard core configuration: it evolves locally towards a configuration in-between the hard and easy cores (Fig. \ref{figDispMaps}). 
This can be seen on the differential displacement map as the length of the horizontal arrow, which separates the downward-pointing triangle defining the hard core from the upward-pointing triangle of the easy core, increases with decreasing atomic fraction of carbon.
One also observes that the centre of the local dislocation density visualised with the Nye tensor is displaced towards the easy core configuration. 
Such a behaviour can be rationalised by the fact that the hard core position is a local energy maximum of the Peierls potential in pure tungsten \cite{Dezerald2014}. 
Thus, the dislocation tends to go back to its more stable easy core configuration when the carbon concentration is not large enough to completely pin it in its unstable hard core configuration. 

\begin{figure}[htbp]
\centering
\includegraphics[width=0.99\linewidth]{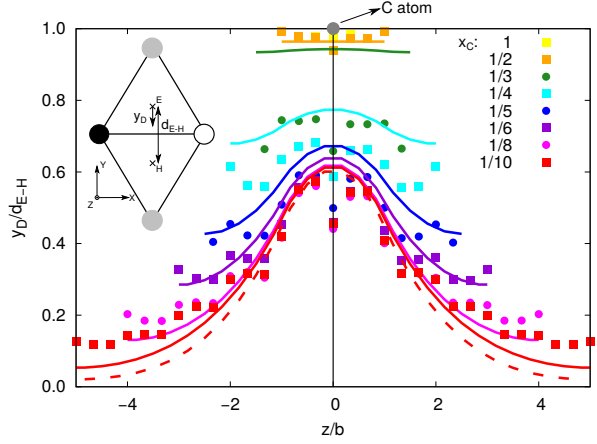}
\caption{Displacement of the dislocation along its line for the different simulation cells investigated in this work.
The symbols correspond to positions deduced from \abinitio{} calculations and the lines to the line tension model.
The dashed line is the profile of an isolated dislocation, thus in the limit of an infinite supercell volume $V$, for $x_{\rm C} = 0.1$.}
\label{figDisloPos}
\end{figure}

Far from the carbon atom along the dislocation line, the dislocation is not pinned anymore in its hard core configuration and tends to fall back in its easy core ground state.
This has been also noted by Bakaev \etal{} \cite{Bakaev2019} who showed for $x_{\rm C}=0.125$
that the carbon atom induces a localised dislocation deflection toward the hard core.
To evidence this core variation along the dislocation line, we have extracted the dislocation position from our \abinitio{} calculations.
This position is obtained in a $1\,b$-thick layer by fitting the differential displacements in the \hkl[111] direction between neighbouring atoms extracted from our \abinitio{} calculations (Fig. \ref{figDispMaps}) to the differential displacements predicted by linear elasticity theory for straight dislocations taking into account periodicity. 
From this fit, we determined the displacement of the dislocation position normalised by its distance between the easy and hard core positions ($y_D/d_{E-H}$) layer-by-layer with a $b/3$ 
discretisation step (Fig. \ref{figDisloPos}).
When the carbon atomic fraction $x_{\rm C}$ is larger than 0.5, the dislocation stays in its hard core position all along its line.
For more dilute carbon concentrations, the pinning of the dislocation is effective only close to the solute and the dislocation tends to go back to its easy core position further.  
The dislocation segment pinned by the carbon atom is located in-between the hard and the easy cores 
with its position converging to the middle point for $x_{\rm C}\leq0.2$.
We also notice that the dislocation does not completely recover its easy core position far from the solute even for the largest supercell ($h=10\,b$).

\begin{figure}[htbp]
\centering
\includegraphics[width=0.99\linewidth]{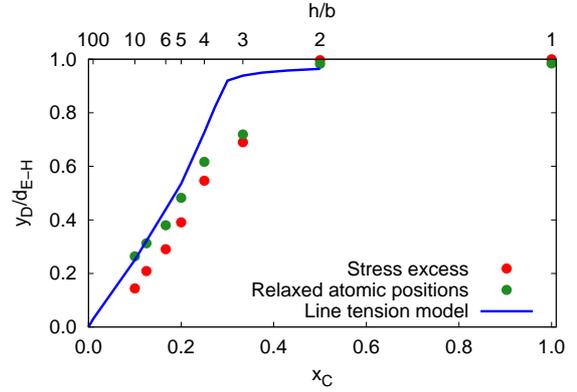}
\caption{Average displacement of the dislocation for the different simulation cells deduced from the stress excess (Eq. \ref{eqAveDisloPos}), from the dislocation profile fitted on relaxed atomic positions (Fig. \ref{figDisloPos}) and compared to the prediction of the line tension model.}
\label{figAveDisloPos}
\end{figure}

To validate the dislocation profile obtained by this fitting procedure, we have compared the average dislocation position corresponding to this layer-by-layer analysis with the average position 
deduced from the stress excess given by our \abinitio{} calculations.
Since fixed periodic boundary conditions are used in our simulations, any change of the dislocation position induces a variation of the homogeneous stress \cite{Chaari2014,Dezerald2016,Rodney2017,Kraych2019}.
The stress variation $\Delta \sigma_{ij}$ observed in our \abinitio{} calculations is linked to the relative positions $x_D$ and $y_D$ of the two dislocations composing the dipole by 
(see reference \cite{Kraych2019} for a derivation of this equation):
\begin{equation}
\label{eqAveDisloPos}
\Delta \sigma_{ij} = \frac{h}{V} C_{ijkl} \left(b_k \Delta A_l - \Delta \Omega_{kl} \right),
\end{equation}
where $\Delta \vec{A} = 2(-\Delta y_D, \Delta x_D, 0)$ is the variation of the dipole cut induced by the motion of the dislocations, $\Delta \Omega$ the variation of the  
dislocation relaxation volume tensor, $C_{ijkl}$ the tungsten elastic constants, and $V$ the simulation cell volume which is kept fixed during relaxation. 
The dislocation positions ($\Delta x_D, \Delta y_D$) are obtained by inverting Eq. (\ref{eqAveDisloPos}) and calculated with respect to the 
dislocation initial positions before atomic relaxation \ie{} the hard core configuration.
Taking now the easy core as the origin, the dislocation positions are $x_D = \Delta x_D$ and $y_D = d_{E-H} - \Delta y_D$.
The average displacement $y_D$ of the dislocations is presented in Fig. \ref{figAveDisloPos} with the average position of the dislocation profile fitted on the differential 
displacements (Fig. \ref{figDisloPos}). 
A good agreement is obtained between both methods, thus giving good confidence in the dislocation profile extracted from our atomic simulations.
Moreover, the average displacement $\Delta x_D$ , also obtained by inverting Eq. (\ref{eqAveDisloPos}), remains inferior to 0.03\,\AA{} for all calculations, thus smaller than the 
displacement $\Delta y_D$ and will be neglected in the line tension model developed in the next section.

\subsection{Line tension model}
\label{S33}

The previous analysis of \abinitio{} calculations shows that the interaction of screw dislocations with carbon is controlled by three physical mechanisms: the pinning by the carbon atom of the dislocation in an unstable position close to the hard core configuration, the dislocation propensity to fall back in its easy core ground state far from the carbon atom, and the limitation of its curvature.
Such an interaction should therefore be well-captured by a line tension model \cite{Rodney2014,Proville2013}.
Assuming a carbon atom at the origin $z=0$, the dislocation position $y(z)$ should minimise the functional:
\begin{multline}
E^{\rm inter}_{\rm D-C}[y(z)] 
= \int{ \left[ E_{\rm D}[y(z)] + \frac{T}{2}\left( \frac{\partial y}{\partial z} \right)^2  \right] \mathrm{d}z} \\
+ \Delta E_{\rm D-C}[y(0)],
\label{eqLT_continu}
\end{multline}
where $E_{\rm D}(y)$ and $T$ are the dislocation line energy and the line tension in pure tungsten respectively, and $\Delta E_{\rm D-C}(y)$ the interaction energy between a carbon atom and an infinite straight dislocation at position $y$.

Considering a periodic array of carbon atoms along the dislocation line with a periodicity length equal to $h$ like in our atomic simulations, we discretise Eq. \ref{eqLT_continu} by decomposing the dislocation line in $N$ segments with their positions defined by $Y_i=y(Z_i)$ and $Z_i=ih/N$.  
These positions should thus minimise the discretised equation:
\begin{multline}
\label{eqModel}
E^{\rm inter}_{\rm D-C}(\lbrace Y_i \rbrace) =  \frac{b}{n} \sum^{N-1}_{i=0} \left[ E_{\rm D}(Y_i) + \frac{T \, n^2}{2 \, b^2}(Y_{i+1}-Y_i)^2\right]  \\
+ \Delta E_{\rm D-C} \left( Y_{\rm C}  \right),
\end{multline}
where we have defined $n=Nb/h$ the number of discrete points per $b$ and $Y_{\rm C} = \frac{1}{2n} \sum_{i=-n}^{n-1}{Y_i}$ the average dislocation position close to the solute.

This discretised equation, without the last contribution describing the dislocation interaction with carbon, has been shown to well reproduce the kink-pair nucleation enthalpy in pure BCC metals,  
including tungsten \cite{Itakura2012,Proville2013,Dezerald2015}.
The same model has also been used sucessfully in  BCC iron to describe interaction of screw dislocations with hydrogen \cite{Itakura2013}.
All the parameters of the line tension model can be derived from \abinitio{} calculations.
In particular, the line tension in tungsten has been already calculated by Dezerald \etal{} \cite{Dezerald2015}, who found $T=3.89$\,eV\,\AA$^{-1}$.

\begin{figure}[htbp]
\centering
\includegraphics[width=0.99\linewidth]{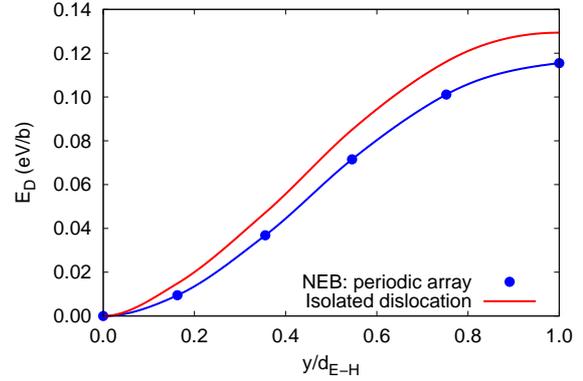}
\caption{Variation of the dislocation self-energy between its easy core and hard core positions obtained for the periodic array used in the \abinitio{} calculations and for an isolated dislocation after subtraction of the variation of the elastic interaction energy.}
\label{figLTMED}
\end{figure}

The variation of the dislocation line energy $E_{\rm D}(y)$ between the easy ($y=0$) and the hard core configuration ($y=d_{\rm E-H}$) has been calculated with the NEB method using the 135-atom  
dislocation supercell with a minimal height $h=b$.  
In this calculation the periodicity vectors were kept fixed and chosen to ensure a zero stress for the hard core configuration, like in the \abinitio{} calculations of \S \ref{S32}.
The dislocation position $y$ has been determined from the excess average stress using Eq. \ref{eqAveDisloPos} for the four intermediate NEB images and the results are shown in Fig. \ref{figLTMED}.  
Due to the small size of the simulation cell and the fixed periodic boundary conditions, part of the obtained energy variation $E_{\rm D}(y)$ corresponds to a variation of the elastic interaction 
between the two dipole dislocations and their periodic images.
We use linear anisotropic elasticity to evaluate this contribution \cite{Rodney2017,Clouet2018} and remove it.
This will allows us to apply the line tension model for an isolated dislocation once we have validated it on the interacting dislocations of our \abinitio{} calculations.
Once this elastic contribution is removed, we obtain an energy difference between the hard and the easy core configuration $\Delta E_{\rm H-E}=129\,\textrm{meV}/b$, in perfect 
agreement with the value previously obtained by just flipping the Burgers vectors of the dislocations (\S \ref{S22}).

\begin{figure}[htbp]
\centering
\includegraphics[width=0.99\linewidth]{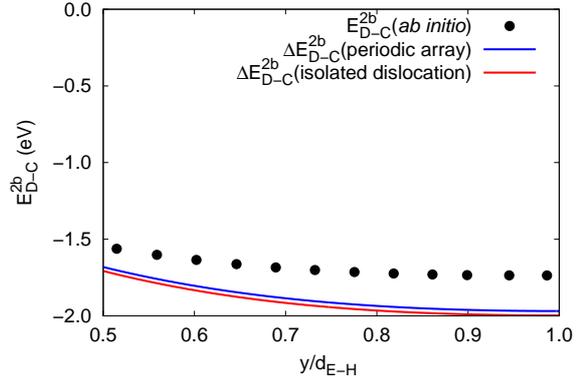}
\caption{Interaction energy between a straight screw dislocation and a carbon atom as a function of the dislocation position $y$ obtained in a simulation cell of height $h=2b$.
The contribution of the dislocation self-energy is then subtracted from this interaction energy 
to obtain the function $\Delta E_{\rm D-C}(y)$.}
\label{figLTMDE_DC}
\end{figure}

The function $\Delta E_{\rm D-C}(y)$ describing the dislocation-carbon interaction energy as a function of the dislocation position $y$ has been obtained with \abinitio{} calculations in a simulation cell of height $h=2b$.
The dislocation position is constrained by fixing, for each \hkl(111) layer, the relative displacements along the \hkl[111] direction of the six tungsten atoms defining the hard core configuration around the carbon atom.
Such a constrained relaxation allows the dilatation caused by the presence of the solute in the dislocation core
but prevents dislocation motion.
The obtained energy $E^{2b}_{\rm D-C}$ shown in Fig. \ref{figLTMDE_DC} also contains the variation of the dislocation self-energy $E_{\rm D}(y)$ which has to be removed so as not to count this contribution twice in Eq. \ref{eqModel} and thus to obtain the function $\Delta E_{\rm D-C}(y)$ used in the line tension model.

Figs. \ref{figEinter}, \ref{figDisloPos} and \ref{figAveDisloPos} compare the carbon-dislocation interaction energy, dislocation profiles and the average dislocation position respectively, obtained directly with \abinitio{} calculations and with this line tension model. 
A good agreement is obtained for all quantities, showing the ability of this simple line tension model to describe the interaction of screw dislocations with carbon.
It is worth pointing out that no parameter in the model has been fitted, with all three quantities entering the model obtained by separate \abinitio{} calculations. 
It can be seen in Figs. \ref{figDisloPos} and \ref{figAveDisloPos} that (i) the dislocation is stabilised in the hard core configuration all along its line with a lower carbon atomic fraction 
using the line tension model ($x_{\rm C}\geq1/3$) than using \abinitio{} calculations ($x_{\rm C}\geq1/2$), and in Fig \ref{figEinter} that (ii) a shift of about 
0.1\,eV is obtained with the line tension model on the dislocation-carbon interaction energy compared to \abinitio{} calculations.
Despite these small differences, the line tension model appears suitable to describe the interaction of screw dislocations with carbon atoms.

Following this validation step, we use the same model to predict this interaction in the dilute regime,
\ie{} for lower carbon concentrations that can not be simulated with \abinitio{} calculations. 
In this dilute regime, we use the functions $E_{\rm D}(y)$ and $\Delta E_{\rm D-C}(y)$, which have been determined for the isolated dislocation (Figs. \ref{figLTMED} and \ref{figLTMDE_DC}).
A converged behaviour is obtained for carbon atomic fractions lower than $x_{\rm C}=0.1$.
The obtained dislocation profile is really close to the one obtained from the same line tension model parametrised for the periodic dislocation array corresponding to the 135-atom supercell ($x_{\rm 
C}=1/10$ in Fig. \ref{figDisloPos}).
Thus, it shows a negligible effect of the periodic boundary conditions enforced in our \abinitio{} calculations.
The converged interaction energy is $E_{\rm D-C}^{\rm inter} =  -1.34$\,eV.


\section{Carbon segregation on dislocation cores}
\label{S4}

We now expand our \abinitio{} calculations to consider other sites that are likely to attract carbon atoms, besides the prismatic sites.
Interaction energies obtained for the most attractive sites are then used to parametrise an Ising model, which we combine with a mean-field approximation to predict carbon equilibrium segregation in the core of screw dislocations \cite{Luthi2019}.

\subsection{Octahedral sites around the reconstructed core}
\label{S41}

\begin{figure}[bth]
\centering\includegraphics[width=0.99\linewidth]{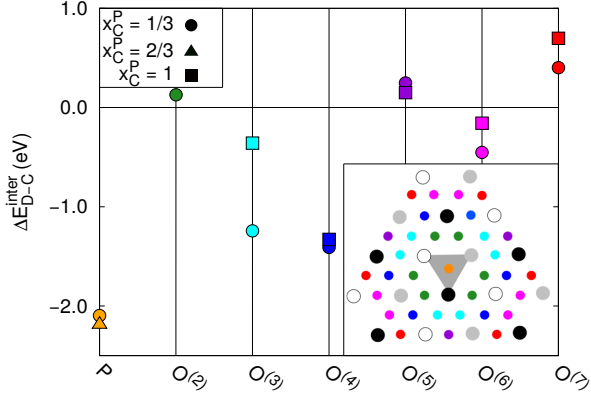}
\caption{Dislocation-carbon incremental interaction energy for the addition of a carbon atom in an empty prismatic ($P$) or an octahedral site $O^{(i)}$ (see inset for a definition of these $O^{(i)}$ sites with the corresponding colour circles
on the \hkl[111] projection of the reconstructed dislocation core). 
The reference for this incremental interaction energy is the reconstructed core with either every third (circles), every two thirds (triangles) or all $P$ sites occupied by a carbon atom (squares).}
\label{figEinterSat}
\end{figure}

We investigate the interaction energy when additional carbon atoms are introduced in octahedral interstitial sites, at the positions of $i-$th nearest neighbours of the reconstructed dislocation core, denoted $O^{(i)}$ with $i$ ranging from 1 to 7.
Starting with a simulation cell of height $3b$ with one prismatic ($P$) site occupied by a carbon atom ($x^{\rm P}_{\rm C}=1/3$), a second carbon atom is added either in an empty $P$ site or in one of the $O^{(i)}$ sites around the prismatic line.
The incremental interaction energy (Eq. \ref{eqEintInc}) corresponding to the insertion of a second carbon atom in a $P$ or $O^{(i)}$ site is represented with circles in Fig. \ref{figEinterSat}, taking as a reference the configuration without the additional carbon atom, \ie{} with every third $P$ sites occupied. 

Since the $O^{(i)}$ sites are situated in $(111)$ planes halfway between the $P$ sites (as shown in Fig. \ref{figVMNT1b}c), the second carbon atom in a $O^{(i)}$ site can be inserted either in 
a (111) plane close to the first solute situated in a $P$ site or close to the two empty $P$ sites.
Hence, we first calculate, for one type of $O^{(i)}$ sites, the variation of interaction energy when the second carbon atom is inserted either close to the first carbon atom or to the two empty prismatic sites. 
The variation of interaction energy is about $2\%$ for $O^{(4)}$, suggesting a very weak interaction between the carbon atoms in a $P$ site and in an $O^{(4)}$ site. This weak interaction will be confirmed later when building the Ising model.
Assuming the same behaviour for others $O^{(i)}$ sites, we arbitrary incorporate the second atom in the $O^{(i)}$ site in the $(111)$ plane close to the occupied prismatic site in the following.

Our DFT calculations for $x^{\rm P}_{\rm C} = 1/3$ lead to stable configurations when the second carbon atom is placed in a octahedral site at a distance equal to or further than second nearest neighbours, with a repulsive interaction for $O^{(2)}$, $O^{(5)}$, and $O^{(7)}$ and an attraction for $O^{(3)}$, $O^{(4)}$ and $O^{(6)}$ (circles in Fig. \ref{figEinterSat}).
The most attractive octahedral positions are $O^{(3)}$ and $O^{(4)}$ sites with $\Delta E^{\rm inter}_{\rm D-C}\simeq-1.5$\,eV.
On the other hand, $O^{(1)}$ insertion sites are unstable: upon relaxation, the inserted carbon atom moves into an empty prismatic site on the dislocation line, resulting in a $x_{\rm C}^{\rm P}=2/3$ occupation of the prismatic line.
This actually leads to the most attractive position, with an incremental interaction energy $\Delta E^{\rm inter}_{\rm D-C}=-2.1$\,eV for the insertion, either directly or through relaxation from an $O^{(1)}$ site, of a carbon atom in such an empty prismatic site.
When a third carbon atom is further added to the latter configuration in the remaining empty prismatic site, resulting in a completely occupied prismatic line, the incremental interaction energy is even slightly more attractive (triangle in Fig. \ref{figEinterSat}), highlighting the persistence of the strong attraction between the dislocation and the prismatic sites.

As the prismatic sites appear more attractive than any octahedral sites, we consider now a completely saturated prismatic line ($x_{\rm C}^{\rm P}=1$), still of height $h=3b$, and study its interaction with a fourth carbon atom located in one of the same octahedral sites (squares in Fig. \ref{figEinterSat}).
The octahedral sites $O^{(3)}$, $O^{(4)}$, and $O^{(6)}$ remain attractive but the interaction is attenuated for $O^{(3)}$ (the interaction energy varies from -1.24 eV to -0.36 eV) and $O^{(6)}$ (it goes from -0.46 eV to -0.16 eV), whereas it remains as strong for the $O^{(4)}$ sites compared to the interaction with the partially populated prismatic 
line.
The octahedral sites $O^{(5)}$ and $O^{(7)}$ remain repulsive and the $O^{(2)}$ sites become unstable with the carbon atom relaxing to the neighbouring $O^{(4)}$ sites.

These calculations show that the prismatic sites are the most attractive insertion sites for carbon with an incremental interaction energy $\Delta E^{\rm inter}_{\rm D-C}\simeq-2.1$\,eV, but that the octahedral $O^{(4)}$ sites of the reconstructed dislocation core are also attractive sites for carbon, with a non negligible interaction energy ($\Delta E^{\rm inter}_{\rm D-C}\simeq-1.5$\,eV).
These results therefore suggest that the dislocation can be decorated by carbon atoms on both $P$ and $O^{(4)}$ sites.

\subsection{Segregation model}
\label{S42}

\subsubsection{Description of interaction energy}
\label{S421}

Based on the above \abinitio{} results, we develop an Ising Hamiltonian, which takes into account both the prismatic  sites and the six different octahedral $O^{(4)}$ sites around the reconstructed dislocation core \cite{Luthi2019}:
\begin{equation}
\label{eqHamil}
H = \sum_{i} \Delta E_i^{seg,0} p_i + \frac{1}{2} \sum_{i,j\neq i} V_{ij} p_i p_j + h\,\Delta E_{H-E},
\end{equation}
with $\Delta E_i^{seg,0}$ the segregation energy on site $i$ ($P$ or $O^{(4)}$) in the dilute limit and $V_{ij}$ the pair interaction between carbon atoms on sites $i$ and $j$. 
The site occupancy $p_i$ is equal to 1 when site $i$ is occupied by a carbon atom and to 0 otherwise.
The first two terms describe the interaction of carbon atoms with the reconstructed dislocation core and the interaction between segregated carbon atoms, whereas the last term, $h\,\Delta E_{H-E}$, corresponds to the energetic cost to transform a dislocation segment of length $h$ from an easy to a hard core. 
For the length $h=3b$ of the \abinitio{} calculations used to parametrise the Ising model, this contribution is equal to 0.387\,eV.

\begin{figure}[bth]
\centering
\includegraphics[width=0.99\linewidth]{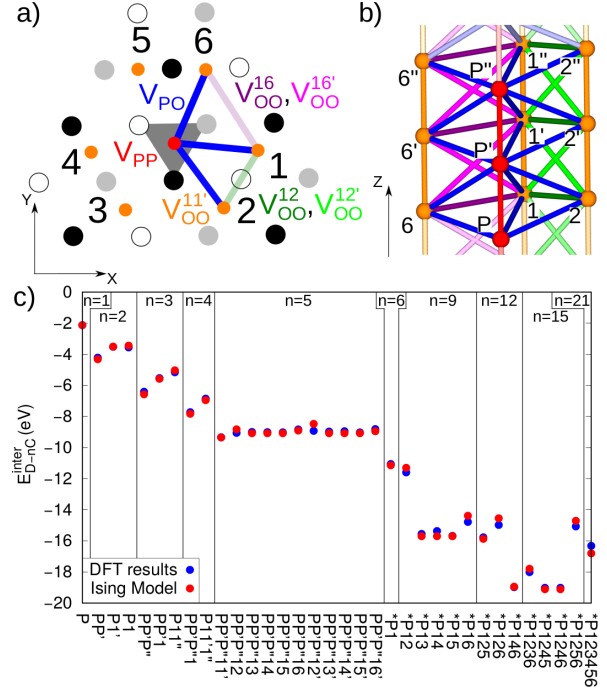}
\caption{Definition and fit of the Ising model. 
(a,b) Schematic representation of the non-null pair interactions: (a) projected in the \hkl(111) plane and (b) along the Burgers vector direction.
(c) Comparison between the interaction energies predicted by the Ising model and obtained by DFT, using a cell of height $3b$ and sorted by the number $n$ of carbon atoms per dislocation.
$P$, $P'$ and $P"$ correspond to the three different $P$ sites along the line, and similarly $i$, $i'$ and $i"$ (with $i$ ranging from 1 to 6) correspond to the three different $O^{(4)}$ sites.
Configurations denoted with a * symbol correspond to atomic columns saturated by carbon, \eg{} configuration named $*P1$ corresponds to carbon atoms located in $P$, $P'$, $P"$, $1$ ,$1'$ and $1"$ positions.}
\label{figEinterDFTIsM}
\end{figure}

We include in Eq. \ref{eqHamil} pair interactions $V_{ij}$ only between carbon atoms which are first nearest neighbours, as it will be proved sufficient.
Hence, a given $P$ site interacts with two $P$ sites (resp. twelve $O^{(4)}$ sites), which are equivalent by symmetry and the corresponding interaction term is denoted $V_{PP}$ (resp. $V_{PO}$).
Concerning the interactions between $O^{(4)}$ sites, the situation is more complex as these sites can be in the same or in different (111) planes (see Fig. \ref{figEinterDFTIsM}b). 
The corresponding interaction terms are denoted $V_{OO}^{ij}$ with $i$ and $j$ the indexes of $O^{(4)}$ sites that can take values from 1 to 6 (see Fig. \ref{figEinterDFTIsM}a).
The interaction between $O^{(4)}$ sites first nearest-neighbours along the dislocation line is denoted $V_{00}^{11'}$.
Then because of symmetry, the interaction $V_{OO}^{12}$ between close neighbours $O^{(4)}$ sites numbered 1 and 2 in a given (111) plane is equivalent to $V_{OO}^{34}$ and $V_{OO}^{56}$ (see Fig. \ref{figEinterDFTIsM}a).
Another interaction, denoted with a prime symbol, corresponds to $O^{(4)}$ sites inside an equivalent pair (1,2), (3,4) or (5,6) but in different (111) planes: $V_{OO}^{12'}$=$V_{OO}^{34'}$=$V_{OO}^{56'}$.
Similarly the equivalent close pairs (1,6), (4,5) and (2,3) involve non-negligible interactions for $O^{(4)}$ sites in a given (111) plane: $V_{OO}^{16} = V_{OO}^{45} = V_{OO}^{23}$, and in different (111) planes: $V_{OO}^{16'} = V_{OO}^{45'} = V_{OO}^{23'}$ (see Fig. \ref{figEinterDFTIsM}b).
All other interactions between $O^{(4)}$ sites involve more distant neighbours and do not need to be considered in the Hamiltonian.

\begin{table}[h]
\centering
\caption{Energy parameters (in eV) of the Ising model (Eq. \ref{eqHamil}).}
\label{tblEseg}
\begin{tabular}{c c}
\hline
$\Delta E_P^{seg,0}$ & $-2.10$ \\ 
$\Delta E_{O^{(4)}}^{seg,0}$  & $-1.38$ \\
$V_{PP}$ & $-0.08$ \\
$V_{PO}$ &  0.06 \\
$V_{OO}^{11'}$ & $-0.27$ \\
$V_{OO}^{12}$ & 0.25 \\
$V_{OO}^{12'}$ & 0.61 \\
$V_{OO}^{16}$ & 0.18 \\
$V_{OO}^{16'}$ & 0.13 \\
\hline
\end{tabular}
\end{table}

We then fit, using the least-squares method, the nine parameters of the Hamiltonian on the \abinitio{} values of dislocation-carbon interaction energies (Eq. \ref{eqEintTot}) calculated for the 34 different configurations listed in Fig. \ref{figEinterDFTIsM}c. 
All these configurations correspond to cells of height $3b$ with $P$ and $O^{(4)}$ columns either partially or completely saturated by carbon.
The resulting set of parameters is given in table \ref{tblEseg} and the interaction energies given by the Ising model are compared to \abinitio{} calculations in Fig. \ref{figEinterDFTIsM}c.
The predictions of the Ising model are very close to \abinitio{} values.
The relative errors between the \abinitio{} results and the Ising model remain inferior to $6\%$ for all configurations, thus showing the ability of this simple Ising model to describe carbon interaction with the screw dislocation.
The obtained parameters (Tab. \ref{tblEseg}) evidence the carbon attraction, already noticed previously, when  solutes are first nearest-neighbours in the direction of the dislocation line.
The attraction is stronger between $O^{(4)}$ sites ($V_{OO}^{11'}=-0.27$\,eV) than between $P$ sites ($V_{PP}=-0.08$\,eV). All the other pair interactions are repulsive.  
However, the repulsion between $P$ and $O^{(4)}$ sites is small: it should not prevent the possibility for carbon atoms to segregate on both sites. 
On the other hand, the repulsion between different $O^{(4)}$ sites is much stronger, probably preventing a possible saturation by carbon atoms of all six $O^{(4)}$ atomic columns around the dislocation core.

\subsubsection{Mean-field approximation}
\label{S422}

In order to determine the equilibrium site concentrations as a function of temperature, we use a mean-field approximation \cite{Duscatelle1991,Ventelon2015,Luthi2019}.
Calling $x_{\rm C}^{\rm P}$ and $x_{\rm C}^{\rm O4}$ the carbon concentrations on $P$ and $O^{(4)}$ sites, the mean-field Hamiltonian $\langle H \rangle$ per unit of $b$ is written as:
\begin{equation}
\label{eqAveHamil}
\begin{aligned}
\langle H \rangle & =  \Delta E_P^{seg,0} x_{\rm C}^{\rm P} + 6\,\Delta E_P^{seg,0} x_{\rm C}^{\rm O4} + V_{PP}\,(x_{\rm C}^{\rm P})^2  \\
& + 12\,V_{PO}\,x_{\rm C}^{\rm P}\,x_{\rm C}^{\rm O4} + 3\,V_{OO}\,(x_{\rm C}^{\rm O4})^2 + b\,\Delta E_{H-E},
\end{aligned}
\end{equation}
with $V_{OO}= 2V_{OO}^{11'}+V_{OO}^{12}+2V_{OO}^{12'}+V_{OO}^{16}+2V_{OO}^{16'}$ the average interaction between $O^{(4)}$ sites.
The occupation of the $P$ and $O^{(4)}$ sites is obtained from the free energy minimisation, using an ideal configurational entropy.
Thus, we obtain:
\begin{equation}
\label{eqConcentration}
\begin{aligned}
\frac{x_{\rm C}^{\rm P}}{1-x_{\rm C}^{\rm P}} = & \frac{x_{\rm C}^{\rm B}}{1-x_{\rm C}^{\rm B}} \\
& \exp{\left(-\frac{\Delta E_P^{seg,0}+2 V_{PP} x_{\rm C}^{\rm P}+12 V_{PO} x_{\rm C}^{\rm O4}}{k_B T} \right)}  \\
\frac{x_{\rm C}^{\rm O4}}{1-x_{\rm C}^{\rm O4}}  = & \frac{x_{\rm C}^{\rm B}}{1-x_{\rm C}^{\rm B}} \\
& \exp{\left( - \frac{\Delta E_O^{seg,0} + V_{OO} x_{\rm C}^{\rm O4}+2 V_{PO} x_{\rm C}^{\rm P}}{k_B T} \right)},
\end{aligned}
\end{equation}
with $k_B$ the Boltzmann constant and $x_{\rm C}^{\rm B}$ the carbon concentration in the matrix.
The latter is linked to the nominal concentration of carbon atoms per tungsten atom, $x_{\rm C}^{\rm nom}$, by matter conservation: $N_B\,x_{\rm C}^{\rm nom}/3 = N_B\,x_{\rm C}^{\rm B} + N_P\,x_{\rm C}^{\rm P} + N_{O4}\,x_{\rm C}^{\rm O4}$, with $N_B$ the number of octahedral sites in the matrix ($N_B=6V/a^3$ for a volume $V$, with $a$ the lattice parameter), $N_P$ and $N_{O4}$ the number of $P$ and $O^{(4)}$ sites ($N_P=\rho V/b$ and $N_{O4}=6N_P$, with $\rho$ the dislocation density).
In this matter conservation equation, we have assumed that $N_B \gg N_P + N_{O4}$ which is verified for any meaningful value of the dislocation density $\rho$.
Incorporating this relation in Eq. \ref{eqConcentration}, which can be solved then self-consistently, we obtain the temperature dependence of the carbon concentrations, $x_{\rm C}^{\rm P}$ and $x_{\rm C}^{\rm O4}$, as a function of the nominal carbon concentration and the dislocation density.

\subsection{Segregation profiles}
\label{S43}

\begin{figure}[htbp]
\centering
\includegraphics[width=0.99\linewidth]{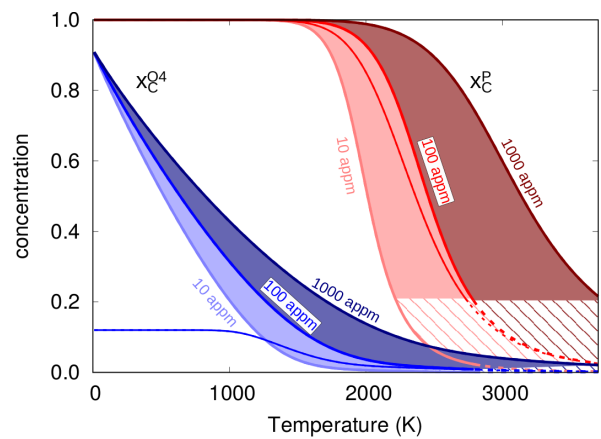}
\caption{Temperature dependence of the carbon concentration segregated in $P$ (in red) and $O^{(4)}$ sites (in blue) for nominal carbon concentrations of 10, 100 and 1000 appm and for dislocation densities of 10$^{12}$ m$^{-2}$ (thick lines) and $10^{15}$ m$^{-2}$ (thin lines).
The dashed lines and dashed areas correspond to the carbon concentration in the different sites beyond the limit of the model.}
\label{figCSeg}
\end{figure}

The carbon concentrations $x_{\rm C}^{\rm P}$ and $x_{\rm C}^{\rm O4}$ as a function of temperature for different nominal concentrations ranging from 10 appm to 1000 appm and for two different dislocation densities ($10^{12}$ m$^{-2}$ and $10^{15}$ m$^{-2}$) are shown in Fig. \ref{figCSeg}.
The $P$ sites are completely saturated by carbon atoms up to at least 1500\,K with a variation similar to the one obtained in the iron-carbon system \cite{Ventelon2015,Luthi2019}.
The exact transition temperature from a completely saturated ($x_{\rm C}^{\rm P}=1$) to an almost empty line ($x_{\rm C}^{\rm P}=0$) increases with the nominal concentration.
Regardless of the nominal concentration, the dislocation density and the temperature, the $O^{(4)}$ sites are less enriched in carbon than the prismatic sites, despite a slower decay with temperature of their concentration.
We note that the concentration of the $O^{(4)}$ sites is not equal to 1 when the temperature tends towards zero even for low dislocation densities and high nominal concentrations of carbon.
This is a consequence of the repulsive interactions existing between the different $O^{(4)}$ sites.
Fig. \ref{figCSeg} also shows that when  dislocation density, $\rho$, increases, \ie{} when the number of possible segregation sites increases, the carbon concentrations in these sites 
decrease, with a much stronger variation for the $O^{(4)}$ than for the $P$ sites.
This variation is actually observed only when $\rho$ becomes higher than $10^{14}$\,m$^{-2}$.
Below this threshold density, segregation profiles do not vary as the number of segregation sites remains low compared to the number of carbon atoms in the solid solution.

The Ising model used to obtain these segregation profiles assume that the dislocation is in a hard core configuration.  
To be valid, enough carbon should be segregated in the prismatic sites to pin the dislocation in this configuration.
In section \ref{S32}, we have shown that the dislocation pinned by carbon remains close to the hard core position all along the line for carbon concentration on prismatic sites $x_{\rm C}^{\rm P}\gtrsim0.2$. 
This necessary minimal concentration on $P$ sites fixes a maximal temperature for the validity of the model. 
Furthermore, the obtained segregation profiles are based only on thermodynamics and therefore assume that carbon can diffuse fast enough to reach equilibrium.
This is not necessary the case at low temperature.  A minimal temperature can be estimated by considering the time necessary for carbon to diffuse over a distance comparable to the average distance between dislocations.
Given the diffusion coefficient of carbon in tungsten \cite{LeClaire1990}, the average time for carbon diffusion over a distance $1/\sqrt{\rho}$ is larger than one hour at 800\,K and 1100\,K when the dislocation density $\rho$ is respectively equal to 10$^{15}$\,m$^{-2}$ and 10$^{12}$\,m$^{-2}$.
Below this temperature which depends on the dislocation density, the diffusion may not be fast enough to reach thermodynamic equilibrium and the corresponding segregation profiles.
For a given carbon nominal concentration, a minimum temperature also exists for which this nominal concentration reaches the solubility limit. 
For a nominal concentration of 1000 and 10\,appm, \ie{} the highest and lowest concentrations considered in Fig. \ref{figCSeg}, this minimum temperature is equal respectively to $\sim$2400\,K and 
$\sim$1300\,K \cite{Goldschmidt1963}.
However above this minimum temperature, corresponding either to a kinetic or a thermodynamic limit, a large range of temperatures remains where screw dislocations are pinned in a hard core 
configuration by an important segregation of carbon in the prismatic sites created by the dislocation lines.  
In this temperature range, there is also a carbon enrichment of the $O^{(4)}$ sites, but this enrichment is not going high enough to completely saturate these sites.

\section{Conclusion}
\label{S5}

Our \abinitio{} calculations evidence a strong attraction between the screw dislocation and the carbon atoms in tungsten, leading to a core reconstruction of the dislocation towards the hard core 
configuration with the solute occupying the prismatic interstitial sites created by the dislocation. 
The reconstruction is perfect for an occupation of prismatic sites $x^P_C$ greater than 0.5, and only partial below.
The interaction of the screw dislocation with carbon atoms leading to its pinning in this hard core configuration is well described by a simple line tension model. 
Such a model does not only allow extrapolation to carbon dilute limit, but could also be used to study the dislocation interaction with a random carbon solid solution.   
Including the work of the applied stress, this model could be further developed to study how carbon affects the nucleation of kink-pairs under stress. 

These prismatic sites in the dislocation core are not the only attractive sites for carbon. 
Our \abinitio{} calculations show that the fourth nearest neighbours octahedral sites of the reconstructed core also attract carbon, with nevertheless a less attractive interaction energy than the prismatic sites
($-1.5$\,eV instead of $-2.1$\,eV).
An Ising model is developed to describe the interaction of the carbon atom with the dislocation pinned in its hard core configuration and also with the other carbon atoms segregated on the line. 
This Ising model combined with a mean-field approximation describes the temperature dependence 
of the carbon concentration in the different segregation sites. 
It predicts that the dislocations remain pinned in this hard core configuration with almost all its prismatic interstitial sites occupied by carbon atoms up to temperatures as high as 2500\,K. 
This carbon segregation on screw dislocations in tungsten is therefore similar to the one already evidenced in iron \cite{Ventelon2015,Luthi2019}, with a different temperature range in agreement 
with the highest melting temperature and the strongest carbon binding in tungsten. 
Furthermore \insitu{} TEM straining experiments in iron-carbon systems \cite{Caillard2015,Caillard2016} have shown that carbon segregation on screw dislocations controls their mobility for 
temperatures high enough to allow for carbon diffusion and lower than $\sim600$\,K where the segregation drops.
Hence, we expect that plasticity in tungsten will be controlled by screw dislocations decorated with carbon atoms up to $\sim2500$\,K.

\vspace{0.5cm}
\linespread{1}
\small

\textbf{Acknowledgments} -
The authors thanks Dr. B. Legrand for fruitful discussions about the thermodynamic model.
They also acknowledge the financial support from the ANR project DeGAS (ANR-16-CE08-0008). 
This work was performed using GENCI-TGCC and GENCI-IDRIS computer centres under Grant No. A0070906821.
It has also been carried out using HPC resources from CINECA computer centre within the framework of the EUROfusion Consortium. 
It has received funding from the Euratom research ans training programme 2014-2018 and 2019-2020 under grant agreement No 633053.
The views and opinions expressed herein do not necessarily reflect those of the European Commission.


\section*{References}
\bibliographystyle{elsarticle-num-names}
\biboptions{sort&compress}
\bibliography{BibActaMat}

\end{document}